\documentclass[aps, oneocolumn, superscriptaddress, showpacs, 12pt]{revtex4}
\usepackage{amssymb}
\usepackage{amsmath}
\usepackage{bm}
\usepackage[dvips]{graphicx}
\usepackage{dcolumn}
\usepackage{longtable}
\usepackage{color}
\usepackage[dvipdfm]{hyperref}

\usepackage{ulem}

\begin{document}

\title{Skyrmion formation in a bulk chiral magnet at zero magnetic field and above room temperature}
\author{K.~Karube}
\altaffiliation{These authors equally contributed to this work}
\affiliation{RIKEN Center for Emergent Matter Science (CEMS), Wako 351-0198, Japan.}
\author{J.S.~White}
\altaffiliation{These authors equally contributed to this work}
\affiliation{Laboratory for Neutron Scattering and Imaging (LNS), Paul Scherrer Institute (PSI),
CH-5232 Villigen, Switzerland.}
\author{D.~Morikawa}
\affiliation{RIKEN Center for Emergent Matter Science (CEMS), Wako 351-0198, Japan.}
\author{M.~Bartkowiak}
\affiliation{Laboratory for Scientific Developments and Novel Materials (LDM), Paul Scherrer Institute (PSI), CH-5232 Villigen, Switzerland}
\author{A.~Kikkawa}
\affiliation{RIKEN Center for Emergent Matter Science (CEMS), Wako 351-0198, Japan.}
\author{Y.~Tokunaga}
\affiliation{Department of Advanced Materials Science, University of Tokyo, Kashiwa 277-8561, Japan.}
\author{T.~Arima}
\affiliation{RIKEN Center for Emergent Matter Science (CEMS), Wako 351-0198, Japan.}
\affiliation{Department of Advanced Materials Science, University of Tokyo, Kashiwa 277-8561, Japan.}
\author{H.M.~R\o nnow}
\affiliation{Laboratory for Quantum Magnetism (LQM), Institute of Physics, \'Ecole Polytechnique F\'ed\'erale de Lausanne (EPFL), CH-1015 Lausanne,
Switzerland.}
\author{Y.~Tokura}
\affiliation{RIKEN Center for Emergent Matter Science (CEMS), Wako 351-0198, Japan.}
\affiliation{Department of Applied Physics, University of Tokyo, Bunkyo-ku 113-8656, Japan.}
\author{Y.~Taguchi}
\affiliation{RIKEN Center for Emergent Matter Science (CEMS), Wako 351-0198, Japan.}

\begin{abstract}
\:
We report that in a $\beta$-Mn-type chiral magnet Co$_9$Zn$_9$Mn$_2$, skyrmions are realized as a metastable state over a wide temperature range, including room temperature, via field-cooling through the thermodynamic equilibrium skyrmion phase that exists below a transition temperature $T_\mathrm{c}$ $\sim$ 400 K. 
The once-created metastable skyrmions survive at zero magnetic field both at and above room temperature. 
Such robust skyrmions in a wide temperature and magnetic field region demonstrate the key role of topology, and provide a significant step toward technological applications of skyrmions in bulk chiral magnets.

\end{abstract}


\maketitle

\newpage

Topological states are in general endowed with robustness against disturbances, such as quenched disorder and thermal agitation, since they cannot be converted into a state with different topology through a continuous deformation process. 
A magnetic skyrmion, a vortex-like spin swirling structure, is one such example of a topological object in magnets, and can be viewed as a particle which is characterized by an integer termed the skyrmion number\cite{Bogdanov,Nagaosa_skyrmion,Muhlbauer,Yu_FeCoSi}. 
Recent extensive studies have clarified that skyrmions are indeed stable to some extent due to their topological nature and exhibit a rich variety of emergent phenomena. In addition, they can be manipulated with a current density of $\sim$ 10$^6$ Am$^{-2}$, which is 5 - 6 orders of magnitude smaller than that required to drive ferromagnetic domain walls, making them promising for spintronics applications\cite{Jonietz,Schulz,Yu_current,Iwasaki,Sampaio}.
Various kinds of magnets have been found to host skyrmions\cite{Muhlbauer,Yu_FeCoSi,Heinze,Romming,Kezsmarki,Ishiwata,Boulle,Woo,Milde,Yu_FeGe,Seki,White,Tokunaga}, and their origins are attributed to several microscopic mechanisms.

One representative mechanism is a competition between a ferromagnetic exchange interaction and a Dzyaloshinskii-Moriya interaction (DMI), the latter of which arises from broken inversion symmetry either at hetero-interfaces\cite{Heinze,Romming,Boulle,Woo,Bode,Hrabec} or in bulk materials\cite{Muhlbauer,Yu_FeCoSi,Kezsmarki,Milde,Yu_FeGe,Seki,White,Tokunaga} with chiral or polar structures. 
For example, the interfacial DMI gives rise to skyrmions in ferromagnetic ultrathin multilayer films at low temperatures\cite{Heinze,Romming} and even at room temperature\cite{Boulle,Woo}.
On the other hand, skyrmions in bulk materials have been investigated in structurally-chiral magnets such as $B$20-type compounds (e.g. MnSi\cite{Muhlbauer}, Fe$_{1-x}$Co$_x$Si\cite{Yu_FeCoSi,Milde}, FeGe\cite{Yu_FeGe}) and Cu$_2$OSeO$_3$\cite{Seki,White}. 
In zero magnetic field, these compounds exhibit a long-period helical magnetic state, described by a single magnetic modulation vector ($q$-vector), as a ground state. 
Slightly below the helimagnetic transition temperature $T_\mathrm{c}$, application of a magnetic field induces the topological transition to a triangular-lattice skyrmion crystal (SkX) state (left of Fig. 1(a)), where the magnetic lattice is often described by triple $q$-vectors that are rotated by 120$^\circ$ with respect to each other and perpendicular to the magnetic field. 
In bulk, the thermodynamical equilibrium SkX state is confined to a narrow temperature and magnetic field region just below $T_\mathrm{c}$, and helical or conical states with zero topological charge are the thermodynamically most stable states at lower temperatures.
However, it is necessary to create skyrmions in a wider temperature and magnetic field region including room temperature and zero magnetic field for both fundamental investigations and practical applications.

Recently, skyrmions have been found as an equilibrium state at and above room temperature in bulk Co-Zn-Mn alloys\cite{Tokunaga} with the $\beta$-Mn-type chiral crystal structure (space group: $P$4$_1$32 or $P$4$_3$32)\cite{Xie,Hori}. 
Mn-free Co$_{10}$Zn$_{10}$ shows a helical state below $T_\textrm{c}$ $\sim$ 480 K, and $T_\textrm{c}$ decreases as the partial substitution with Mn proceeds. 
In Co$_8$Zn$_8$Mn$_4$ ($T_\textrm{c}$ $\sim$ 300 K), a triangular-lattice SkX appears as an equilibrium state around 290 K under a magnetic field.
Although the temperature is around room temperature, this equilibrium phase exists only in a narrow temperature and magnetic field region similar to the other materials. On the other hand, it was found that a supercooled metastable SkX state is realized by field-cooling (FC) with a moderate rate of $\sim$ 1 K min$^{-1}$. Once created, the metastable SkX survives over a wide temperature region, and the SkX coordination transforms into a square-like one (right of Fig. 1(a)) at low temperatures in the metastable state\cite{Karube,Morikawa}.
Despite these advances, skyrmions prevailing in a wide range of temperature and magnetic field, including above room temperature and zero field, remain to be demonstrated in bulk chiral magnets. 
In the present study, we focused on Co$_9$Zn$_9$Mn$_2$ ($T_\textrm{c}$ $\sim$ 400 K) and performed small angle neutron scattering (SANS) and ac susceptibility measurements on a bulk sample, and Lorentz transmission electron microscopy (LTEM) measurement on a thin-plate specimen. 
As summarized in the state diagram in Fig. 1(b), we have succeeded in observing a highly robust metastable SkX for a wide temperature and magnetic field range including room temperature and zero magnetic field both in the bulk and thin-plate samples. 

Single-crystalline bulk samples of Co$_9$Zn$_9$Mn$_2$ were grown by the Bridgman method as reported elsewhere\cite{Tokunaga,Karube}. 
SANS measurements were performed with neutron wavelength 10 $\mathrm{\AA}$ using the instrument SANS-I at the Paul Scherrer Institute, Switzerland.
Ac susceptibility measurements were performed by a superconducting quantum interference device magnetometer (MPMS3, Quantum Design) equipped with an ac susceptibility measurement option. 
LTEM observations were performed with a transmission electron microscope (JEM-2100F, JEOL). 
For all the measurements in this paper, a magnetic field was applied along [110] direction of the crystal.
Details of sample preparation and measurements are described in Supplementary Material. 

First, we show the results of SANS measurements in Fig. 2.
We observed the hallmark 6-spot pattern originating from the triangular-lattice SkX at 390 K and 0.04 T as the equilibrium state. 
One of the triple $q$-vectors of the triangular-lattice SkX points to the vertical [001] direction as in the case of Co$_8$Zn$_8$Mn$_4$\cite{Tokunaga,Karube}. 
From the observed value of $q$ $\sim$ 0.0485 nm$^{-1}$ the lattice constant of triangular-lattice SkX is calculated as $a_\mathrm{sk}$ = $4\pi/\sqrt{3}* q^{-1}$ $\sim$ 150 nm.
In the FC process from the equilibrium SkX state (384 K $\leq$ $T$ $\leq$ 396 K), the 6-spot pattern persists down to 300 K (although the spots become broader), clearly demonstrating the realization of the metastable SkX around room temperature. 
In contrast, after zero-field cooling (ZFC) to 300 K and subsequent field-increasing processes (Supplementary Figs. S3 and S4), the 6-spot pattern is never observed for any temperature and magnetic field condition.

When, after the FC process, the magnetic field is decreased to zero while keeping the temperature at 300 K, the 6-spot pattern gradually changes to a 4-spot pattern. Here, the horizontal 2 spots are much broader than the vertical 2 spots, indicating that the broad horizontal spot is a superposition of the two original, now weaker spots, and a new, relatively stronger spot in between. 
From the rocking-scan data (Supplementary Fig. S5), we found that these 2 broad horizontal spots lie on the plane perpendicular to the incident neutron beam and are clearly distinct from the helical $q$-vectors aligned with the [100] and [010] directions that are out of the plane. 
The increased intensity at the horizontal positions is also distinct from what is expected for the other helical state (or partially merged skyrmions as reported in Fe$_{1-x}$Co$_x$Si\cite{Milde}) with a preferred $q$-vector aligned with the vertical [001] direction (as found at 390 K, 0 T). 
Therefore, we conclude that the broad 4-spot pattern at zero magnetic field after the FC represents a mixture of the original triangular-lattice SkX and a square-like lattice SkX, as illustrated at the left and right of Fig. 1(a), respectively. 

As shown in Supplementary Fig. S6, when the magnetic field is increased back to 0.04 T and then the temperature is increased while keeping the magnetic field at 0.04 T, the broad 4-spot pattern finally changes back to a broad 6-spot pattern at 375 K, well below the temperature region of the equilibrium SkX phase. This indicates that the metastable SkX undergoes the reversible transformation between triangular lattice and square-like lattice accompanied by a broad coexistence region and a large hysteresis, and demonstrates clearly that the metastable skyrmions persist at 300 K and zero field. 
The transition to a square-lattice SkX at low magnetic fields in the metastable state has been also observed in Co$_8$Zn$_8$Mn$_4$\cite{Karube} and MnSi\cite{Nakajima}. 
This common SkX transformation can be qualitatively explained in terms of the overlapping of adjacent skyrmions due to the reduced peripheral region at low magnetic fields, as theoretically demonstrated\cite{Lin}. 

In the subsequent warming process (as indicated by a red arrow in Fig. 2(a)) from 300 K while keeping the magnetic field zero, the metastable SkX with the 4-spot pattern survives up to at least 375 K, which is also clearly distinct from the SANS patterns in the ZFC process (Supplementary Figs. S3 and S5), and finally changes to the equilibrium helical state at 390 K. 
The temperature variation of the SANS intensity originating from metastable skyrmions is displayed in Fig. 2(c). 

We also confirmed that the metastable SkX at 300 K survives in the very wide magnetic field region up to the conical-ferromagnetic phase boundary and even down to the negative field region as shown in Fig. 2(d) (see also Supplementary Fig. S4 and S5).

Next, we show the results of ac susceptibility measurements in Fig. 3. 
In the field dependence at 300 K shown in Fig. 3(b), a large and asymmetric (with respect to the sign of the field) hysteresis is observed between $-$0.09 T and 0.23 T after a FC process (FC1) via the equilibrium SkX phase. On the other hand, the ZFC process and another FC process (FC2) that bypasses the equilibrium SkX phase show smaller and symmetric hystereses that are related to changes between multi-domain helical states and the conical state. 
These indicate that the asymmetric hysteresis observed only after FC1 stems from the supercooled metastable SkX, which is consistent with the SANS data shown in Fig. 2(d).
Similar field sweepings after FC via the equilibrium SkX phase were performed at several temperatures (Fig. 3(c)), clearly indicating the realization of the metastable SkX in a wide temperature and magnetic field region including room temperature and zero magnetic field.

Figure 3(d) shows the temporal dependence of the normalized ac susceptibility at several temperatures after FC via the equilibrium SkX phase. 
The observed relaxation time from the metastable SkX to the equilibrium conical state (lifetime of the metastable SkX) is quite long even at 380 K near the equilibrium SkX phase boundary. 
The relaxation time further increases at lower temperatures and finally the relaxation behavior is hardly observable at 350 K. 
In Fig. 3(e), the estimated relaxation time $\tau$ is plotted against temperature and fitted to a modified Arrhenius law\cite{Oike}. 
The activation energy of the decay rate $\tau^{-1}$ reaches as high as $10^4$ K even at 350 K, well above room temperature. Remarkably, this protection energy scale for the metastable skyrmions is much higher than the other magnetic interaction energies, such as the ferromagnetic interaction and the DMI (several hundred K and several K, respectively, in the present case).
The obtained coefficient of relaxation time $\tau_0$ $\sim$ 47 s is 5 orders of magnitude longer than that in MnSi ($\tau_0$ $\sim$ 10$^{-4}$ s)\cite{Oike}. 
Since the critical cooling rate for quenching an equilibrium phase to lower temperatures is inversely correlated to $\tau_0$, 
the moderately slow cooling rate ($\sim$ 1 K min$^{-1}$) is sufficient to quench the equilibrium SkX state in the present case.
A plausible origin of the long lifetime of metastable skyrmions in Co$_9$Zn$_9$Mn$_2$ is the pinning effect caused by random site occupancies\cite{Hori} (8$c$ site is mainly occupied by Co while 12$d$ site is occupied by random mixture of Co, Zn, and Mn) and a larger skyrmion size (i.e., the number of spins involved in one skyrmion) of 150 nm as compared to 18 nm in MnSi. 
A similar metastable SkX accessible by a moderate FC has been reported in Fe$_{1-x}$Co$_x$Si alloys\cite{Milde,Munzer} with random site occupancy and a large skyrmion size (90 nm at $x$ = 0.5), in which the origin may be the same as the present case.

We further confirmed the formation of skyrmions by real-space observation using LTEM on a (110) thin-plate sample with the thickness of $\sim$ 150 nm. 
As shown in Fig. 4(b), we observed the triangular-lattice SkX state at approximately 370 K and 0.07 T as the equilibrium state. The SkX state persists down to 290 K as the metastable state in a FC process.
The metastable SkX at 290 K is hardly influenced upon decreasing the magnetic field, and remains triangularly coordinated at zero magnetic field. 
The characteristic swirling of the magnetic moments in a skyrmion is clearly shown by the angular color coding in Fig. 4(c) deduced by a transport-of-intensity equation analysis\cite{Ishizuka}.
The metastable SkX survives even at negative fields at 290 K (Supplementary Fig. S7).
In the warming process, while keeping magnetic field zero, the triangular-lattice SkX survives up to at least 350 K and finally changes to the equilibrium helical state at 370 K. 

In comparison with the SANS result, while the metastable SkX is commonly observed irrespective of the sample thickness, both the preferred $q$-vector alignments of helical states and the lattice forms of skyrmions at zero field are different between bulk and thin-plate samples. 
These differences arise presumably from shape anisotropy, namely, significant dipole interactions in the thin-plate sample. In this case the dipole interaction effectively works as a repulsive force among skyrmions, tending to preserve the close packed triangular lattice, and thereby preventing the SkX lattice transformation in the case of the thin-plate sample. Magnetocrystalline anisotropy tends to favor magnetic moment directions along [100] and its equivalent\cite{Xie}, thus giving rise to the preferred $q$-vector direction along [001] in the bulk. For the case of the (110) thin-plate sample, shape anisotropy towards the in-plane magnetization coincides with the magnetocrystalline anisotropy for $q$ $\parallel$ [$-$110], but not for $q$ $\parallel$ [001], thus leading to $q$ $\parallel$ [$-$110] as observed in Fig. 4(b).

In conclusion, we have demonstrated the realization of highly robust metastable skyrmions in a wide temperature and field range, including room temperature and zero magnetic field, in Co$_9$Zn$_9$Mn$_2$ by field-cooling from an equilibrium skyrmion phase. 
The robustness of skyrmions in such a wide parameter space is characteristic for topological spin textures accompanied with both quenched random disorder and a slowly varying nature of the magnetic moment within a single skyrmion. 
The robust metastable skyrmions arising from the Dzyaloshinskii-Moriya interaction associated with the chiral structure of the bulk thus present as highly suitable for investigating both the fundamental properties of topological spin textures, and also practical applications of skyrmions.

\begin{acknowledgments}
We are grateful to  N. Nagaosa, W. Koshibae, X. Z. Yu, F. Kagawa, T. Nakajima and H. Furukawa for fruitful discussions. 
This work was supported by JSPS Grant-in-Aids for Scientific Research (S) No.24224009, the Swiss National Science Foundation (SNF) Sinergia network `NanoSkyrmionics', the SNF projects 153451 and 166298, and the Europea Research Council project CONQUEST.
\end{acknowledgments}

\bibliographystyle{apsrev}

%%%%%%%%%%%%%%%%%%  FIG1  %%%%%%%%%%%%%%
\begin{figure}[tbp]
\begin{center}
\includegraphics[width=10cm]{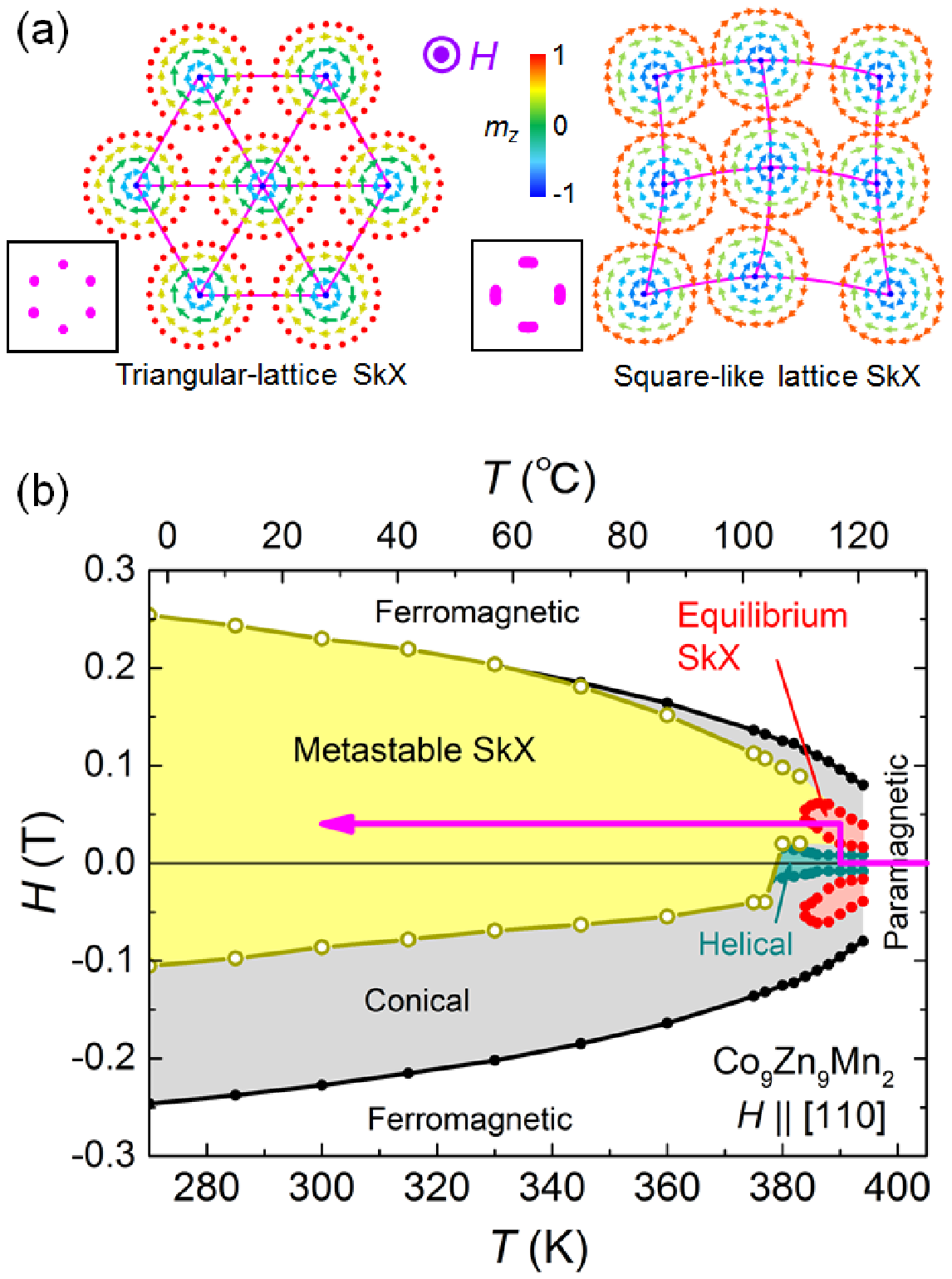}
\caption{(a) Schematic figures of a triangular-lattice SkX and a square-like lattice SkX in real space. 
Magnetic moments are indicated by arrows, and their $z$ components ($m_z$) parallel to the magnetic field ($H$) are represented by colors. 
The corresponding reciprocal space (SANS) patterns are also indicated in the boxes at left sides of the real space configurations.
(b) Temperature - magnetic field state diagram of bulk Co$_9$Zn$_9$Mn$_2$.
The phase boundaries of helical multi-domain (dark green area), thermodynamical equilibrium SkX (red area), conical (gray area), and induced-ferromagnetic states (white area below $T_\mathrm{c}$ $\sim$ 396 K), all indicated with closed symbols, are determined by isothermal ac susceptibility measurements in field-increasing runs after ZFC from 400 K to the measurement temperature (see Supplementary Fig. S2). 
The boundaries of the metastable SkX state (yellow area) denoted with open symbols are determined from isothermal ac susceptibility measurements done as field-sweeping runs from 0.04 T to $\pm$0.3 T after FC via the equilibrium SkX phase as described by the pink arrow (see Fig. 3(c)). }
\end{center}
\end{figure}
%%%%%%%%%%%%%%%%%%%  FIG1   %%%%%%%%%%%%%

%%%%%%%%%%%%%%%%%%  FIG2  %%%%%%%%%%%%%%
\begin{figure}[tbp]
\begin{center}
\includegraphics[width=11cm]{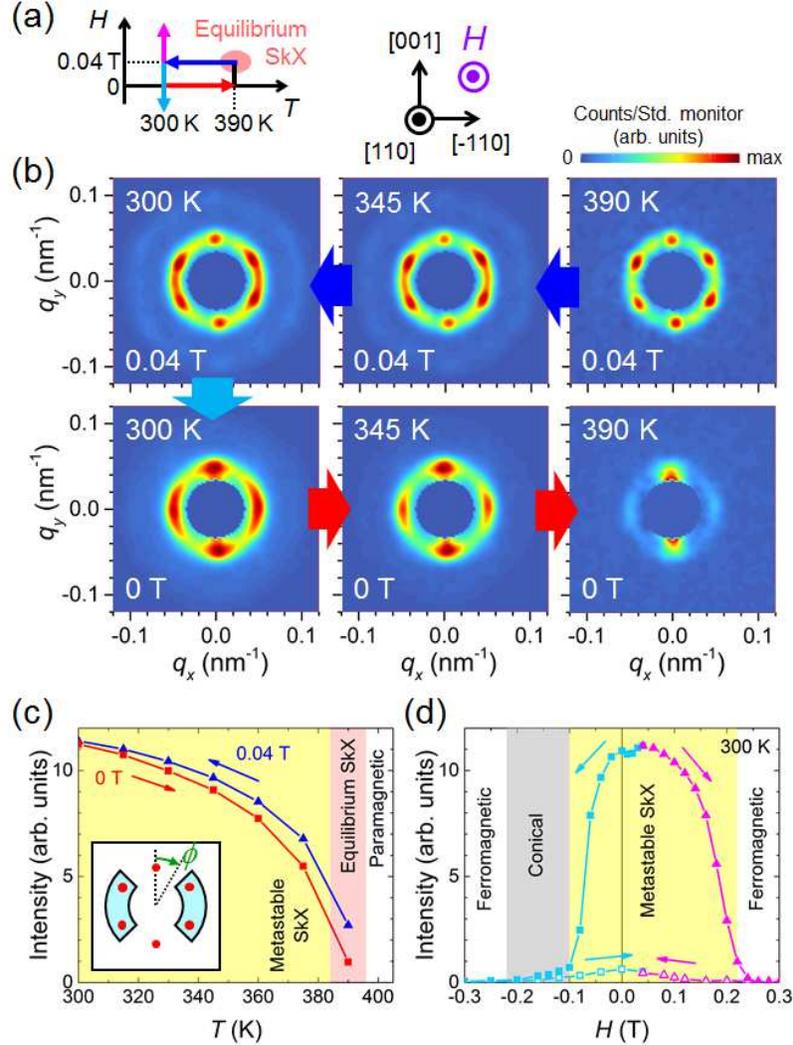}
\caption{(a) Schematic illustration of the measurement process. 
The colors of the arrows correspond to the colors of arrows or data points in panels (b)-(d).
(b) SANS images in the FC process and in the subsequent zero-field warming process.
The intensity scale of the color plots varies between each panel. 
The crystal orientation and field direction are indicated at the top.
(c) Temperature dependence of the SANS intensity integrated over the the azimuthal angle region at $\phi = 90\pm45^\circ$ and $270\pm45^\circ$  (light blue region in the inset). 
Here, $\phi$ is defined as the clockwise azimuthal angle from the vertical [001] direction.
(d) Field dependence of the SANS intensity at 300 K after the FC, integrated over the same azimuthal angle area as in panel (c) (see also Supplementary Fig. S4). 
The intensity in returning processes from $\pm$0.3 T back to 0.04 T is denoted by the open symbols with the same colors. }
\end{center}
\end{figure}
%%%%%%%%%%%%%%%%%%%  FIG2   %%%%%%%%%%%%%

%%%%%%%%%%%%%%%%%%  FIG3  %%%%%%%%%%%%%%
\begin{figure}[tbp]
\begin{center}
\includegraphics[width=9cm]{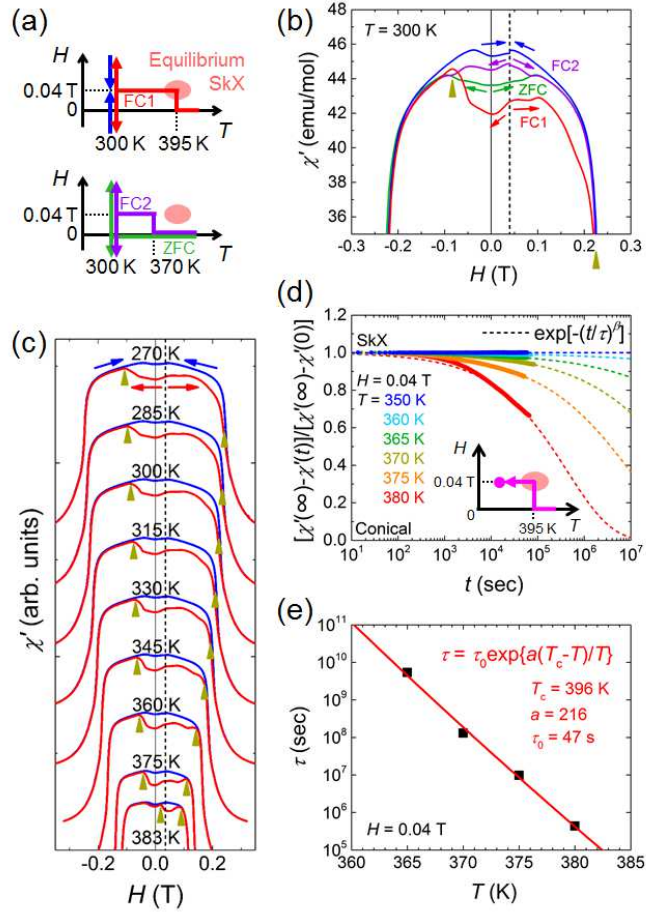}
\caption{(a) Schematic illustrations of the measurement processes shown in panel (b). 
(b) Field dependence of the real part of the ac susceptibility ($\chi^\prime$) at 300 K. 
The results of the field sweepings (not shown) from $\pm$0.3 T back to 0.04 T (0 T) after FC2 (ZFC) are almost the same as that after FC1.
The yellow triangles denote the boundaries of the metastable SkX state. 
(c) Field dependences of $\chi^\prime$ at several different temperatures after the FC process via the equilibrium state. 
The yellow triangles denote the boundaries of the metastable SkX that are plotted in the state diagram in Fig. 1(b).
(d) Temporal dependences of the normalized $\chi^\prime (t)$ at 0.04 T and several temperatures after the FC process via the equilibrium state. 
$\chi^\prime (0)$ is an initial value (metastable SkX state), and $\chi^\prime (\infty)$ is a fully relaxed value (equilibrium conical state) which is assumed to be the value of $\chi^\prime$ at 0.04 T in the field-decreasing run from 0.3 T.
The data points are fitted to the stretched exponential functions (dotted lines). 
(e) Temperature dependence of the relaxation time $\tau$ determined by the fits in panel (d).
The experimental data (black symbols) are fitted to a modified Arrhenius law (red line)\cite{Oike}. 
}
\end{center}
\end{figure}
%%%%%%%%%%%%%%%%%%%  FIG3   %%%%%%%%%%%%%

%%%%%%%%%%%%%%%%%%  FIG4  %%%%%%%%%%%%%%
\begin{figure}[tbp]
\begin{center}
\includegraphics[width=12cm]{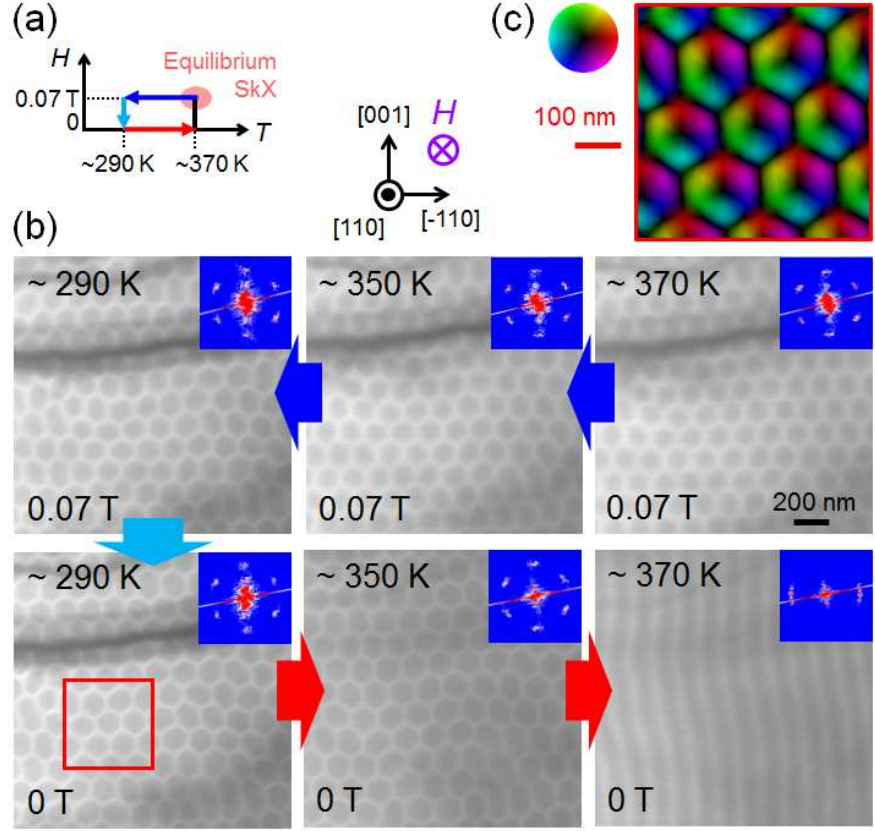}
\caption{(a) Schematic illustration of the measurement process shown in panel (b). 
(b) Over-focused LTEM images on the (110) plane in the FC process and in the subsequent zero-field warming process. 
The crystal orientation and field direction are indicated at the top.
These LTEM images were taken for the nearly same sample position. 
The insets of the respective panels show the fast Fourier-transformed patterns of the LTEM images for a wider sample area.
(c) Distribution of in-plane magnetic moments for the area indicated with red square in panel (b) at 290 K and zero magnetic field after the FC as deduced from a transport-of-intensity equation analysis of over-focused and under-focused LTEM images. 
The color wheel represents the direction and magnitude of the in-plane magnetization.}
\end{center}
\end{figure}
%%%%%%%%%%%%%%%%%%%  FIG4   %%%%%%%%%%%%%

\end{document}

\title{Supplementary Material for \\ ``Skyrmion formation in a bulk chiral magnet at zero magnetic field and above room temperature"}
\author{K.~Karube}
\altaffiliation{These authors equally contributed to this work}
\affiliation{RIKEN Center for Emergent Matter Science (CEMS), Wako 351-0198, Japan.}
\author{J.S.~White}
\altaffiliation{These authors equally contributed to this work}
\affiliation{Laboratory for Neutron Scattering and Imaging (LNS), Paul Scherrer Institute (PSI),
CH-5232 Villigen, Switzerland.}
\author{D.~Morikawa}
\affiliation{RIKEN Center for Emergent Matter Science (CEMS), Wako 351-0198, Japan.}
\author{M.~Bartkowiak}
\affiliation{Laboratory for Scientific Developments and Novel Materials (LDM), Paul Scherrer Institute (PSI), CH-5232 Villigen, Switzerland}
\author{A.~Kikkawa}
\affiliation{RIKEN Center for Emergent Matter Science (CEMS), Wako 351-0198, Japan.}
\author{Y.~Tokunaga}
\affiliation{Department of Advanced Materials Science, University of Tokyo, Kashiwa 277-8561, Japan.}
\author{T.~Arima}
\affiliation{RIKEN Center for Emergent Matter Science (CEMS), Wako 351-0198, Japan.}
\affiliation{Department of Advanced Materials Science, University of Tokyo, Kashiwa 277-8561, Japan.}
\author{H.M.~R\o nnow}
\affiliation{Laboratory for Quantum Magnetism (LQM), Institute of Physics, \'Ecole Polytechnique F\'ed\'erale de Lausanne (EPFL), CH-1015 Lausanne,
Switzerland.}
\author{Y.~Tokura}
\affiliation{RIKEN Center for Emergent Matter Science (CEMS), Wako 351-0198, Japan.}
\affiliation{Department of Applied Physics, University of Tokyo, Bunkyo-ku 113-8656, Japan.}
\author{Y.~Taguchi}
\affiliation{RIKEN Center for Emergent Matter Science (CEMS), Wako 351-0198, Japan.}
\maketitle

\newpage

\subsection*{Experimental details}

\subsubsection*{Sample preparation}
An ingot of Co$_9$Zn$_9$Mn$_2$ composed of several single-crystalline grains was grown by the Bridgman method. 
A single-crystalline grain was cut along the (110), ($-$110) and (001) planes with rectangular shape for small angle neutron scattering (SANS) and ac susceptibility measurements as shown in Fig. S1(a) and (b), respectively. 
For the Lorentz transmission electron microscopy (LTEM) measurement, a tiny thin-plate sample with (110) face and the thickness of approximately 150 nm was prepared from a single-crystalline bulk sample by a focused ion beam (FIB) of Ga. 

\subsubsection*{SANS measurement}
SANS measurements were performed using the instrument \textit{SANS-I} at the Paul Scherrer Institute, Switzerland.
High-temperature measurements above room temperature were performed using an oven stick designed for SANS experiments. 
The neutron beam was collimated over a length of 18 m before the sample. The scattered neutrons were counted by a 2D position-sensitive multi-detector located 20 m behind the sample. The neutron wavelength was selected to be 10 $\mathrm{\AA}$ with a 10\% full width at half maximum (FWHM) spread.
The mounted single-crystalline sample was installed into a horizontal field cryomagnet so that the field direction was parallel to the [110] direction. 
Maintaining the $H$ $\parallel$ [110] geometry, the rocking angle $\omega$ between neutron beam and the field direction is scanned in 2$^\circ$ steps around the $H$ $\parallel$ beam ($\omega$ = 0) configuration, by rotating (`rocking') the cryomagnet together with the sample around the vertical [001] direction (Fig. S5(b)).
The observed widths of the peaks (the FWHM) in scattering intensity versus $\omega$ were always broader than $\sim$ 5$^\circ$. Therefore, the SANS images shown in the main text are obtained by summing multiple SANS measurements taken over the range of $-$16$^\circ$ $\leq$ $\omega$ $\leq$ 10$^\circ$ in order to observe both the intensity and positions of all of the Bragg spots in a single image. 
The scanning condition of temperature and field in the FC process is as follows: (i) degaussing at 420 K, (ii) ZFC to 390 K, (iii) applying field of  0.04 T, (iv) FC to measurement temperatures with a cooling rate of $\sim$ $-$2.0 K min$^{-1}$ across the phase boundary (384 K) of the equilibrium SkX phase. The measurement time at each stabilized temperature and field with the above rocking-scan was about 30 min. 

\subsubsection*{Ac susceptibility measurement} 
Ac susceptibility measurements were performed by a superconducting quantum interference device magnetometer (MPMS3, Quantum Design) equipped with an ac susceptibility measurement option. 
The static magnetic field and the ac excitation magnetic field (1 Oe, 193 Hz) were applied along the [$-$110] direction.
Due to the difference in shape between the samples used in the ac susceptibility and the SANS measurements, their demagnetization factors are different. 
In order to eliminate this difference, the field values for the ac susceptibility measurements are calibrated as $H_\mathrm{c} = 2.0*H$ (see Fig. S2(c)).
For all the figures related to ac susceptibility in the main text and Supplementary Material, the calibrated value $H_\mathrm{c}$ is used with the notation of $H$.
The scanning condition of temperature and field in the FC process is as follows: (i) degaussing at 400 K, (ii) ZFC to  395 K, (iii) applying field of 0.04 T, (iv) FC to measurement temperatures with the cooling rate with $\sim$ $-$10 K min$^{-1}$ across the phase boundary (384 K) of the equilibrium SkX phase. 
Magnetic field was swept with a rate of 0.02 T s$^{-1}$ and stabilized at each measurement point every 4 mT. 
The measurement time for each field was about 23 s. 

\subsubsection*{LTEM measurement} 
LTEM observations were performed with a transmission electron microscope (JEM-2100F) at an acceleration voltage of 200 kV.
Temperatures above room temperature were controlled by using a single-tilt heating holder. 
The temperatures presented in the main text and Supplementary Material correspond to the holder temperature, therefore maybe slightly different from the actual sample temperatures up to by approximately $\pm$10 K.
A magnetic field was applied perpendicular to the (110) plate and its magnitude was controlled by tuning the electric current of the objective lens.
Due to the thin-plate sample shape ($\sim$ 150 nm - thick), its demagnetization factor and thus magnetic field scale are larger than those in bulk samples.
The defocus values of LTEM images were set to be +288 $\mu$m at 370 K, +192 $\mu$m at 350 K, and +96 $\mu$m at 290 K, respectively.
The lateral magnetization distribution was obtained by the transport-of-intensity equation analysis of the over-focused and under-focused LTEM images.

\subsection*{Identification of equilibrium SkX state}
The equilibrium skyrmion crystal (SkX) state was identified by SANS and ac susceptibility measurements in the field-sweeping process at 390 K after zero-field cooling (ZFC) (Fig. S2). 
As shown in the SANS images of Fig. S2(b), a helical state with 2 spots along the vertical [001] direction appears at 0 T. 
The weak side intensities are due to finite contributions from the tails of the other 2 helical domains with broad Bragg spots aligned with [100] and [010] directions, respectively, as demonstrated by the rocking curve analysis. 
With increasing magnetic field, a triangular-lattice SkX state with a 6-spot pattern appears at 0.04 T as an equilibrium state. As shown in Fig. S2(d), the integrated SANS intensity originating from the equilibrium SkX shows large values in the range of 0.02 T $\leq$ $H$ $\leq$ 0.06 T, where the ac susceptibility shows a dip (Fig. S2(c)). Above 0.06 T, the triangular-lattice SkX state changes to a conical state, in which the $q$-vector is parallel to the magnetic field, and no Bragg spot is observed in the present SANS configuration. 
Similar field-sweeping measurements for ac susceptibility were performed at different temperatures, and the phase boundaries of the equilibrium SkX state on $T$-$H$ plane was determined (Fig. 1(b) in the main text).

\subsection*{ZFC process in the SANS measurement}
The temperature variation of the SANS images in a ZFC process from 420 K to 300 K is shown in Fig. S3(b). The helical multi-domain state with 2 stronger vertical spots and 2 weaker horizontal spots was observed down to 300 K. These SANS images are compared with those in the zero-field warming process after the field cooling (FC) via the equilibrium SkX phase displayed in Fig. S3(c). The relative intensities of the 2 horizontal spots to those of the 2 vertical spots are clearly distinct between these two processes, indicating that the metastable SkX state survives in zero magnetic field above room temperature. This point is further discussed in terms of the rocking curve analysis.

\subsection*{Field-sweeping processes in the SANS measurement}
The field variations of SANS images at 300 K after a FC at 0.04 T are shown in Fig. S4(b). In the positive field direction, the 6-spot pattern originating from the triangular-lattice SkX persists up to 0.14 T. In the negative field direction, the intensities of the 4 side spots become weaker, and instead the intensities of the 2 vertical spots become slightly stronger; consequently the total SANS pattern changes gradually to a 4-spot pattern with broad side spots. This variation of the SANS pattern indicates the gradual transformation of the SkX from the triangular lattice to a square-like lattice with a broad region of coexistence. Further sweeping the field in the negative direction, the metastable SkX changes to a conical state and no Bragg spot is observed at $-$0.14 T. The SANS intensities originating from the metastable SkX state in these processes are plotted in Fig. 2(d) in the main text.
As shown in Fig. S4(c), in the field sweeping at 300 K after a ZFC, the helical multi-domain state with the stronger 2 vertical spots and weaker 2 horizontal spots remains at 0.04 T and changes to a conical state at 0.14 T without any trace of SkX states. 

\subsection*{Rocking curves in the SANS measurement}
The temperature and field variations of rocking curves of the SANS intensities are shown in Figs. S5(c) and (d), respectively.
The relation between the field direction, crystal axes, incident neutron beam and rocking angle $\omega$ is illustrated in Fig. S5(b).
In the FC process (blue triangle symbols in Fig. S5(c)), the rocking curves show a maximum at $\omega$ = 0$^\circ$ down to 300 K since $q$-vectors of the metastable triangular-lattice SkX state are parallel to the (110) plane and perpendicular to the field. In the subsequent zero-field warming process (red square symbols), the rocking curves also show peaks at $\omega$ $\sim$ 0$^\circ$ up to 375 K although the intensities are slightly lower than those in the FC process, demonstrating that horizontal $q$-vectors of the SANS images in Fig. S3(c) remain parallel to the (110) plane, as expected for a SkX state. On the other hand, the rocking curves in the ZFC process (green circle symbols) show broad minimum around $\omega$ $\sim$ 0$^\circ$ down to 300 K since the $q$-vectors of two of the helical multi-domains are pointing along the [100] and [010] directions that are out of the (110) plane. Thus, the 2 broad side spots of the 4-spot pattern observed in the zero-field warming process after the FC process arise not from a helical state but from a coexistence of a square-like lattice SkX and the original triangular-lattice SkX.
In the field sweeping to negative fields after the FC (light blue square symbols in Fig. S5(d)), the peak structure of the rocking curves is preserved down to $-$0.04 T. These are also different from the valley-shaped rocking curves in the field sweeping after the ZFC process (orange circle symbols), demonstrating the survival of the metastable SkX down to negative fields.

\subsection*{Returning processes in the SANS measurement}
Figure S6(b) shows the temperature and field variation of the SANS images in the returning process (yellow arrow in Fig. S6(a)) after the FC to 300 K and the subsequent field sweeping to zero field. 
When the magnetic field is returned to 0.04 T at 300 K, the broad 4-spot pattern (coexistence of a triangular-lattice SkX and a square-like lattice SkX) persists. With increasing the temperature while keeping the magnetic field at 0.04 T, the broad 4-spot pattern remains up to higher temperatures and finally changes to a broad 6-spot pattern (disordered triangular-lattice SkX) at 375 K.
It is noted that 375 K is well below the temperature range of the equilibrium SkX phase (384 K $\leq$ $T$ $\leq$ 396 K) as confirmed by the field sweeping at 375 K after a ZFC, in which  no trace of SkX states was observed  (Fig. S6(c)).
Therefore, the metastable SkX shows a reversible structural transition between a triangular lattice and a square-like lattice with a broad coexistence region and a large hysteresis in the above temperature and field sweeping procedures after the FC. This provides unambiguous evidence that the metastable skyrmions survive at 300 K and zero field.

\subsection*{Field-sweeping process in the LTEM measurement}
To further demonstrate the survival of metastable skyrmions in a negative field region, we show the field variation of real-space images by Lorentz transmission electron microscopy (LTEM) on a thin-plate sample in Fig. S7(b).
As discussed in the main text, the metastable triangular-lattice SkX at 290 K created by a FC via the equilibrium SkX phase persists at zero magnetic field.
Upon increasing the magnitude of the field in the negative direction, although some skyrmions merge with each other and the total topological charge decreases, a number of skyrmions remain and keep the original triangular lattice at $-$0.049 T, clearly demonstrating the survival of a significant metastable skyrmion density in a negative field region.
The observed positional dependence may arise from inhomogeneity of effective magnetic fields (due to a demagnetization effect) or random pinning for the spin textures. 
Further increasing the field in the negative direction, the number of skyrmions that merge with each other increases while some topological structures still remain at $-$0.078 T, and finally almost all the skyrmions are destroyed and change to a helical state.

\newpage

%%%%%%%%%%%%%%%%%%  FIG  %%%%%%%%%%%%%%
\begin{figure}[tbp]
\begin{center}
\includegraphics[width=8cm]{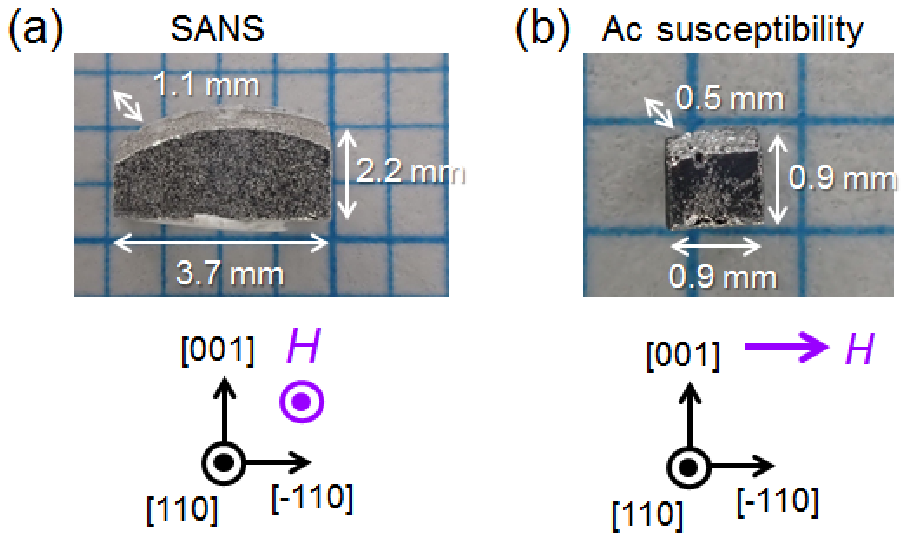}
\caption{Pictures of single-crystalline bulk samples of Co$_9$Zn$_9$Mn$_2$ used for (a) SANS and (b) ac susceptibility measurements. 
Directions of the crystal axes and applied magnetic field ($H$) are also indicated.}
\end{center}
\end{figure}
%%%%%%%%%%%%%%%%%%%  FIG   %%%%%%%%%%%%%

\newpage

%%%%%%%%%%%%%%%%%%  FIG  %%%%%%%%%%%%%%
\begin{figure*}[tbp]
\begin{center}
\includegraphics[width=13cm]{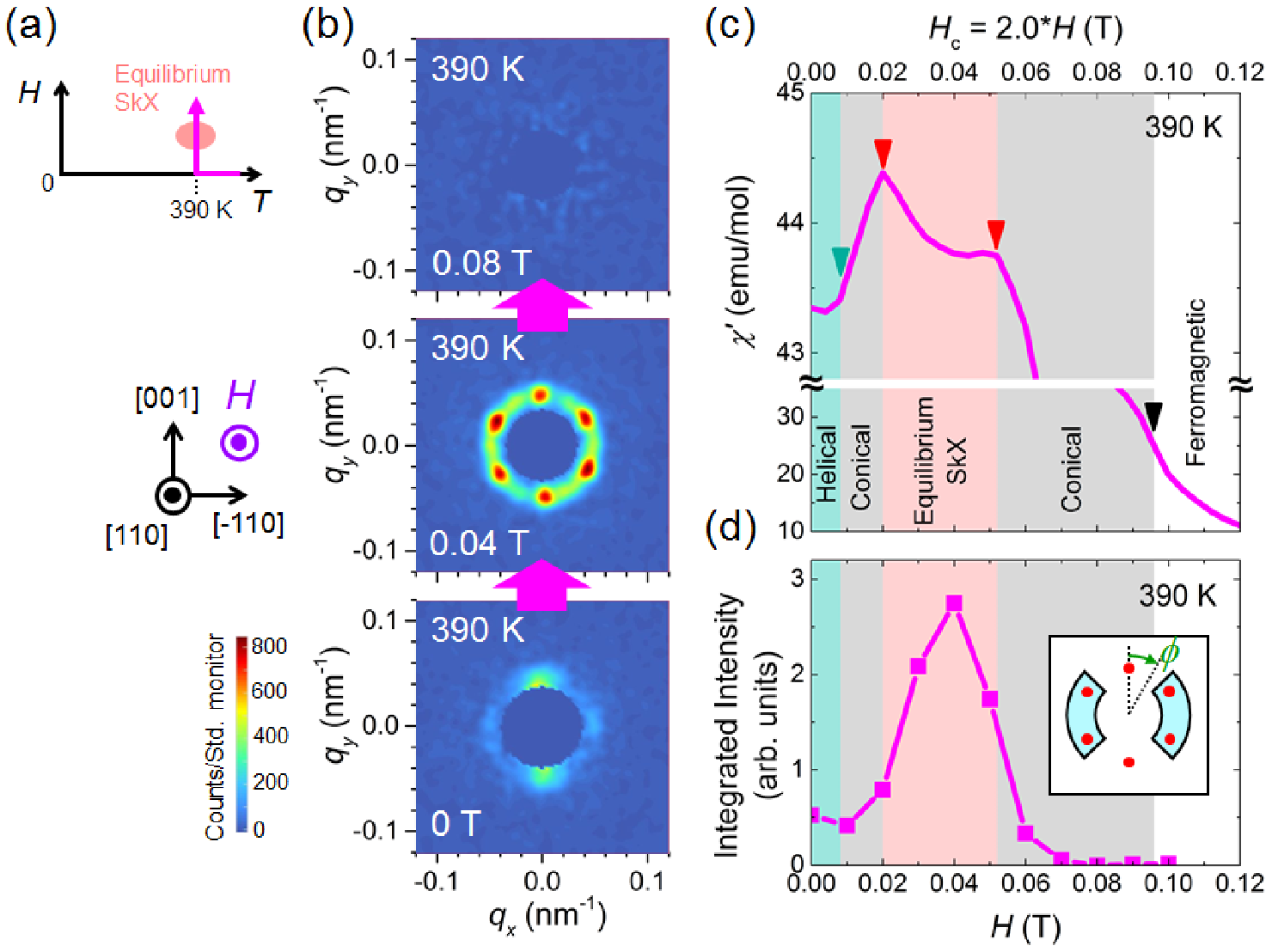}
\caption{(a) Schematic illustration of the measurement process. 
(b) SANS images at 390 K and selected fields. 
The intensity scale of the color plot is fixed for the three panels.
(c) Field dependence of the real-part of the ac susceptibility ($\chi^\prime$). 
Since the demagnetization factor is different to that in the SANS measurement, the field values are calibrated as $H_\mathrm{c} = 2.0*H$.
The phase boundary (dark green triangle) between helical and conical states was determined as the inflection point of $\chi^\prime$. 
The phase boundaries (red triangles) between equilibrium SkX and conical states were determined as the peak positions of $\chi^\prime$.
The phase boundary (black triangle) between conical and induced-ferromagnetic states was determined as the inflection point of $\chi^\prime$. 
These boundaries and similar data points at different temperatures are plotted as closed symbols in the state diagram of Fig. 1(b) in the main text. 
(d) Field dependence of the SANS intensity integrated over the azimuthal angle region at $\phi = 90\pm45^\circ$ and $270\pm45^\circ$ (light blue region in the inset).
}
\end{center}
\end{figure*}
%%%%%%%%%%%%%%%%%%%  FIG   %%%%%%%%%%%%%

\newpage

%%%%%%%%%%%%%%%%%%  FIG  %%%%%%%%%%%%%%
\begin{figure}[tbp]
\begin{center}
\includegraphics[width=15cm]{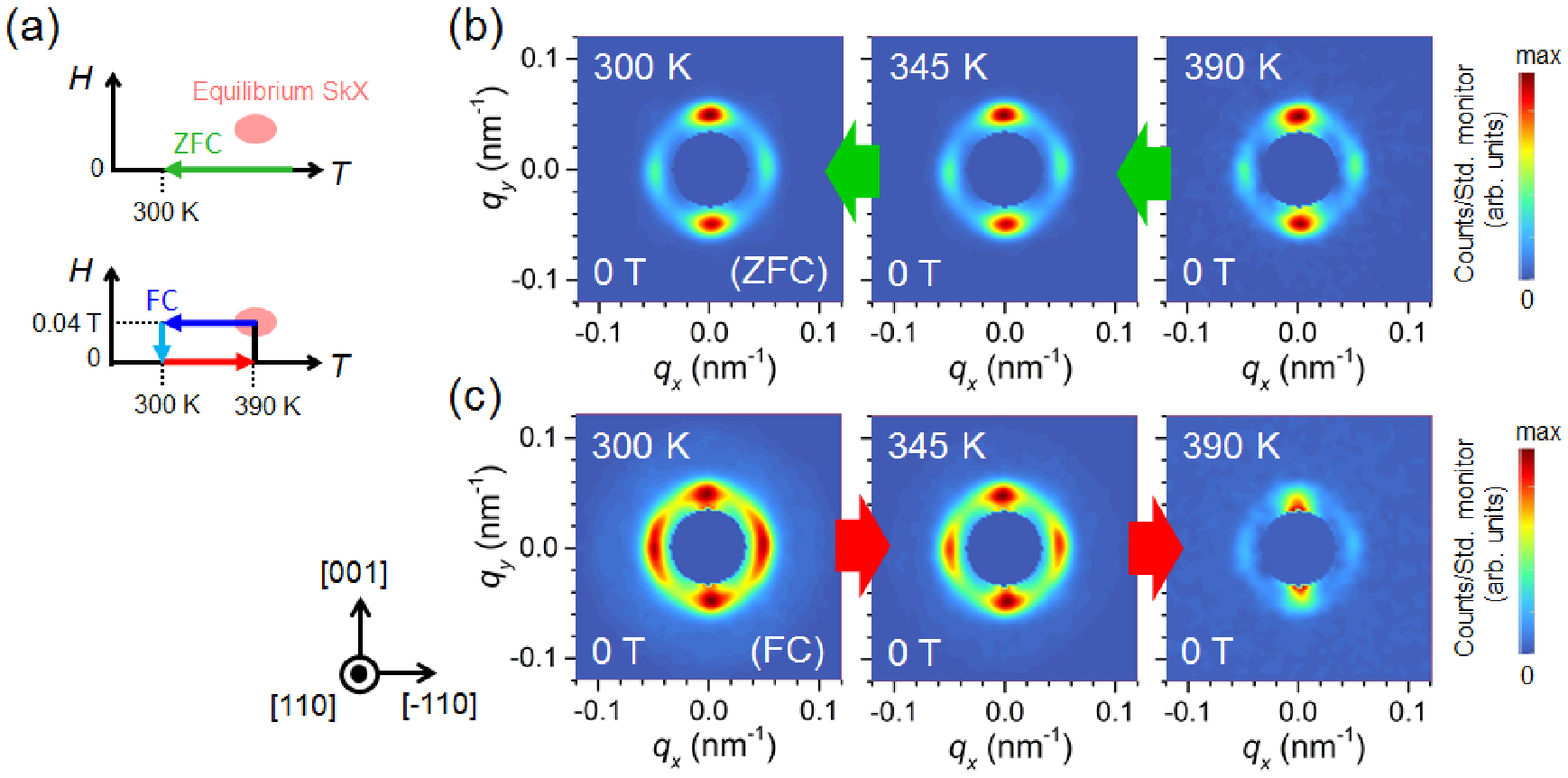}
\caption{(a) Schematic illustrations of the measurement processes for panels (b) and (c). 
(b) SANS images at 390 K, 345 K, and 300 K in a ZFC process from 420 K (indicated by green arrows).
The intensity scale of the color plot varies between each panel. 
(c) SANS images at 300 K, 345 K, and 390 K in the zero-field warming process after a FC (indicated by red arrows), which are the same as Fig. 2(b) in the main text, are displayed again for a better comparison with panel (b).
}
\end{center}
\end{figure}
%%%%%%%%%%%%%%%%%%%  FIG   %%%%%%%%%%%%%

\newpage

%%%%%%%%%%%%%%%%%%  FIG  %%%%%%%%%%%%%%
\begin{figure}[tbp]
\begin{center}
\includegraphics[width=16cm]{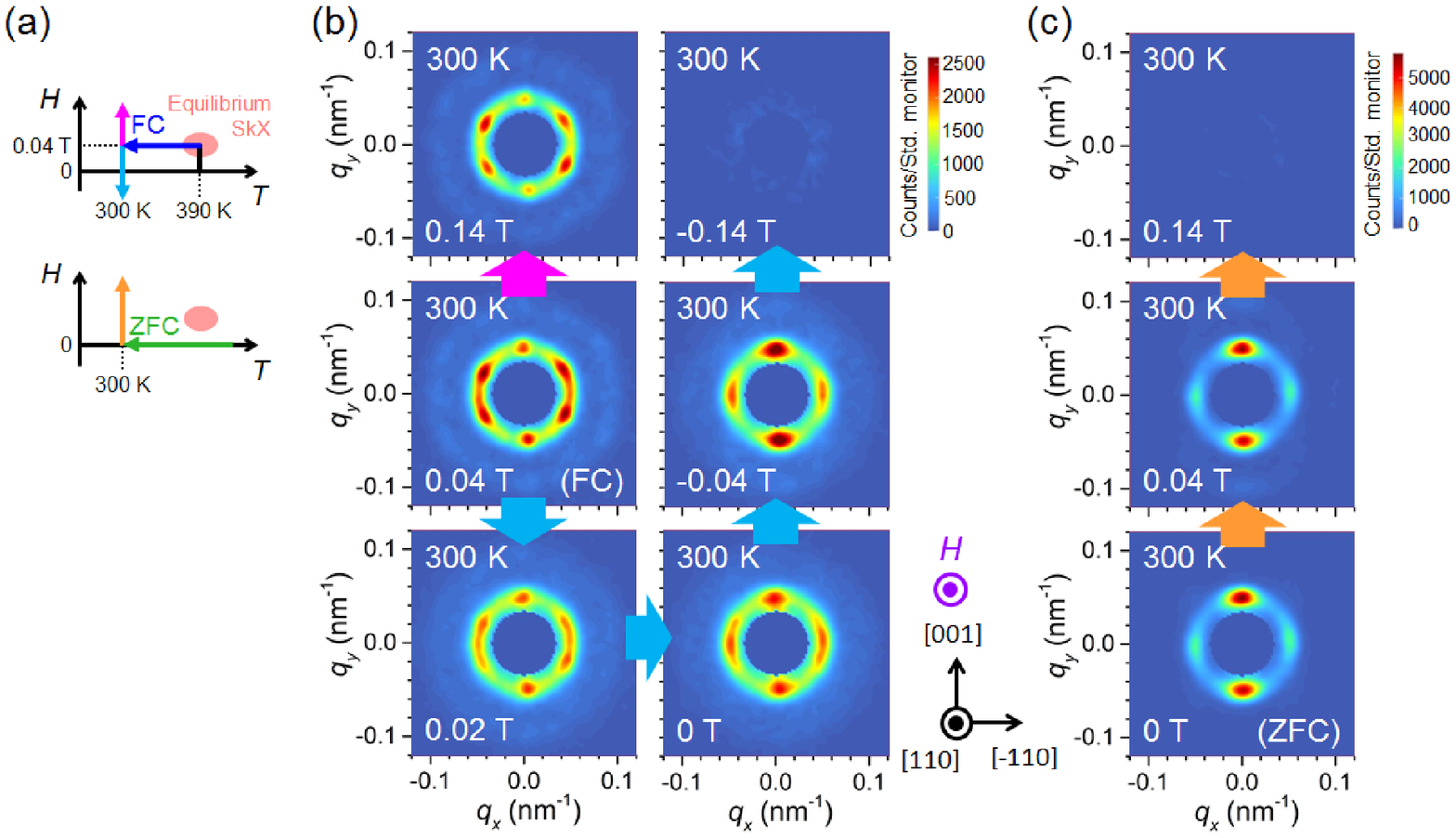}
\caption{(a) Schematic illustrations of the measurement processes for panels (b) and (c). 
(b) SANS images at 300 K and selected magnetic fields in the field-sweeping processes after the FC. 
Field sweepings from 0.04 T to 0.3 T and to $-$0.3 T are indicated by a pink arrow and light blue arrows, respectively. 
The intensity scale of the color plot is fixed for the six panels.
(c) SANS images at 300 K and selected magnetic fields in the field-sweeping process after a ZFC (indicated by orange arrows). 
The intensity scale of the color plot is fixed for the three panels.
}
\end{center}
\end{figure}
%%%%%%%%%%%%%%%%%%%  FIG   %%%%%%%%%%%%%

\newpage

%%%%%%%%%%%%%%%%%%  FIG  %%%%%%%%%%%%%%
\begin{figure}[tbp]
\begin{center}
\includegraphics[width=12.5cm]{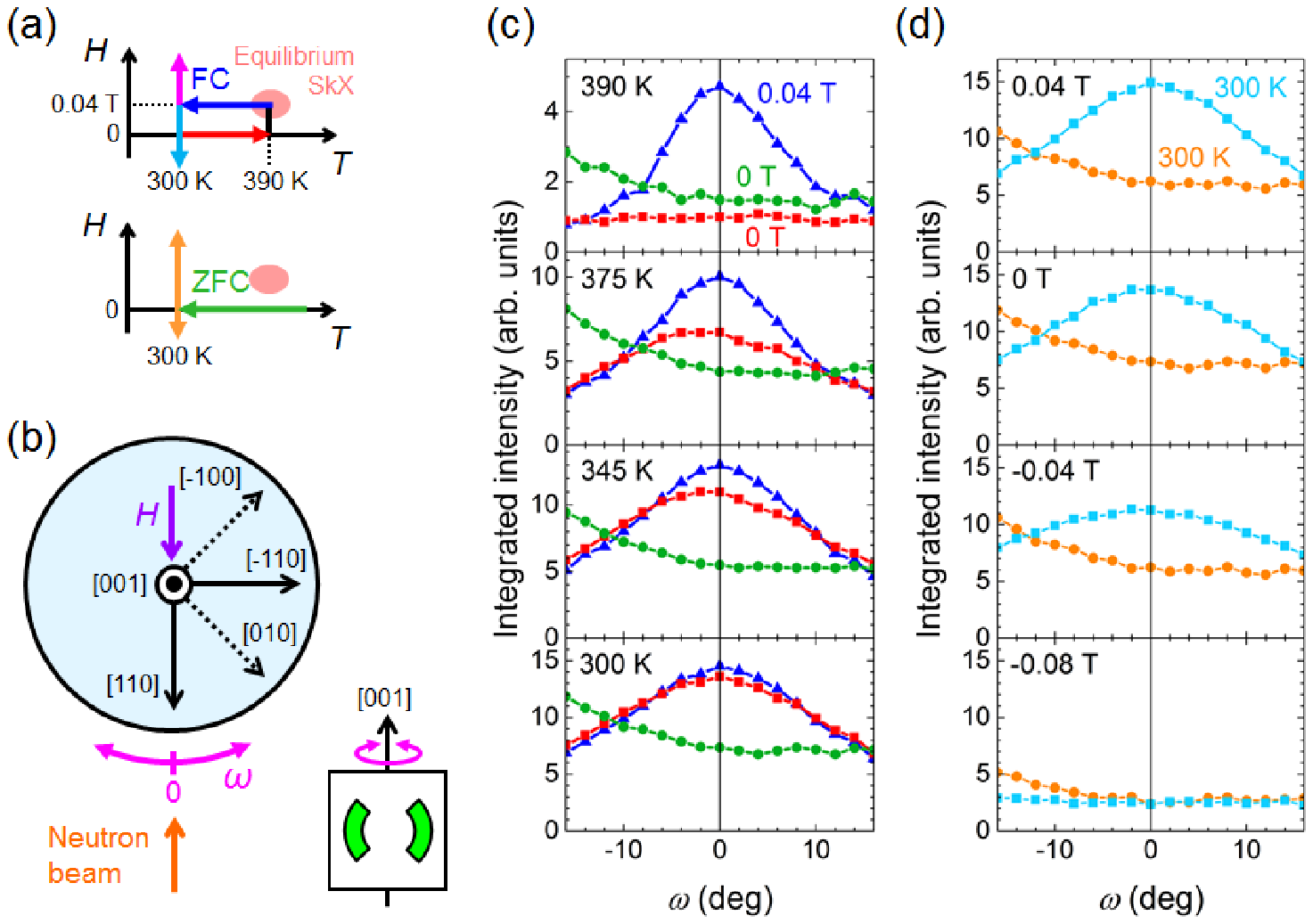}
\caption{(a) Schematic illustrations of the measurement processes. 
(b) The top view (along the vertical [001] direction) of the experimental configuration is schematically depicted. 
Maintaining the $H$ $\parallel$ [110] geometry, the angle $\omega$ between the neutron beam and the field direction is varied (`rocked') around the vertical [001] direction.
The green area of the inset shows the azimuthal angle region ($\phi = 90\pm45^\circ$ and $270\pm45^\circ$) over which the integrated intensities are evaluated and plotted in panels (c) and (d).
(c) Rocking angle dependences of the integrated intensities at selected fields and temperatures.
The data for the FC process from 390 K to 300 K and for the subsequent zero-field warming process from 300 K to 390 K are indicated with blue triangle and red square symbols, respectively.
The data for the ZFC process is plotted with green circle symbols.
(d) Rocking angle dependences of the integrated intensities at 300 K and selected magnetic fields.
The data for the field sweeping from 0.04 T to $-$0.08 T after the FC is displayed with light blue square symbols. 
The data for the field sweepings from 0 T to 0.04 T and to $-$0.08 T after the ZFC are shown with orange circle symbols. 
}
\end{center}
\end{figure}
%%%%%%%%%%%%%%%%%%%  FIG   %%%%%%%%%%%%%

\newpage

%%%%%%%%%%%%%%%%%%  FIG  %%%%%%%%%%%%%%
\begin{figure}[tbp]
\begin{center}
\includegraphics[width=16cm]{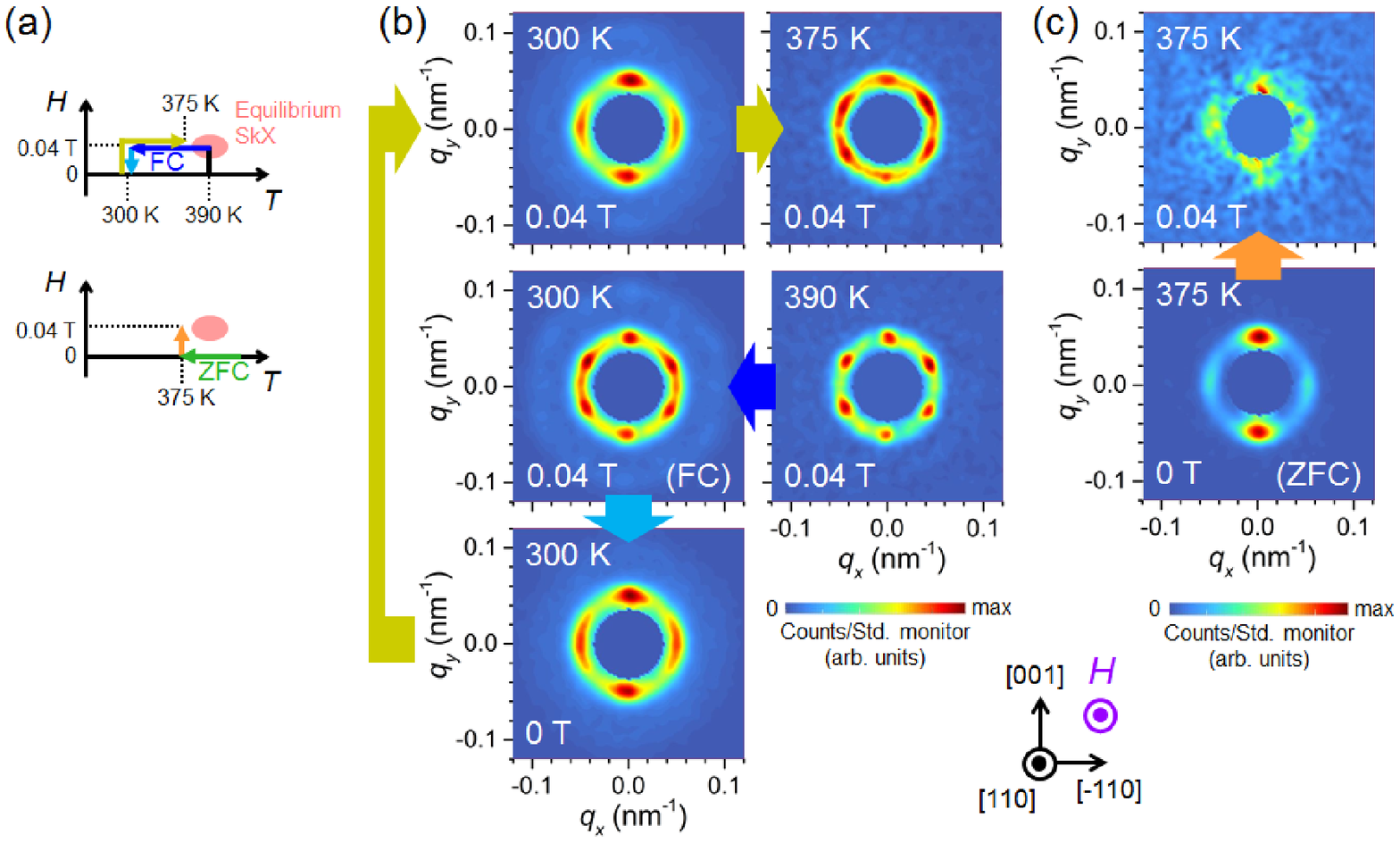}
\caption{(a) Schematic illustrations of the measurement processes for panels (b) and (c). 
(b) SANS images at selected temperatures and magnetic fields in the FC to 300 K (blue arrow), the subsequent field-sweeping to zero field (light blue arrow) and the subsequent returning processes (yellow arrows).
The intensity scale of the color plot varies between the five panels. 
(c) SANS images at 375 K and at 0 T and 0.04 T in the field-sweeping process after a ZFC (orange arrow). 
The intensity scale of the color plot varies between the two panels. 
}
\end{center}
\end{figure}
%%%%%%%%%%%%%%%%%%%  FIG   %%%%%%%%%%%%%

\newpage

%%%%%%%%%%%%%%%%%%  FIG  %%%%%%%%%%%%%%
\begin{figure}[tbp]
\begin{center}
\includegraphics[width=12cm]{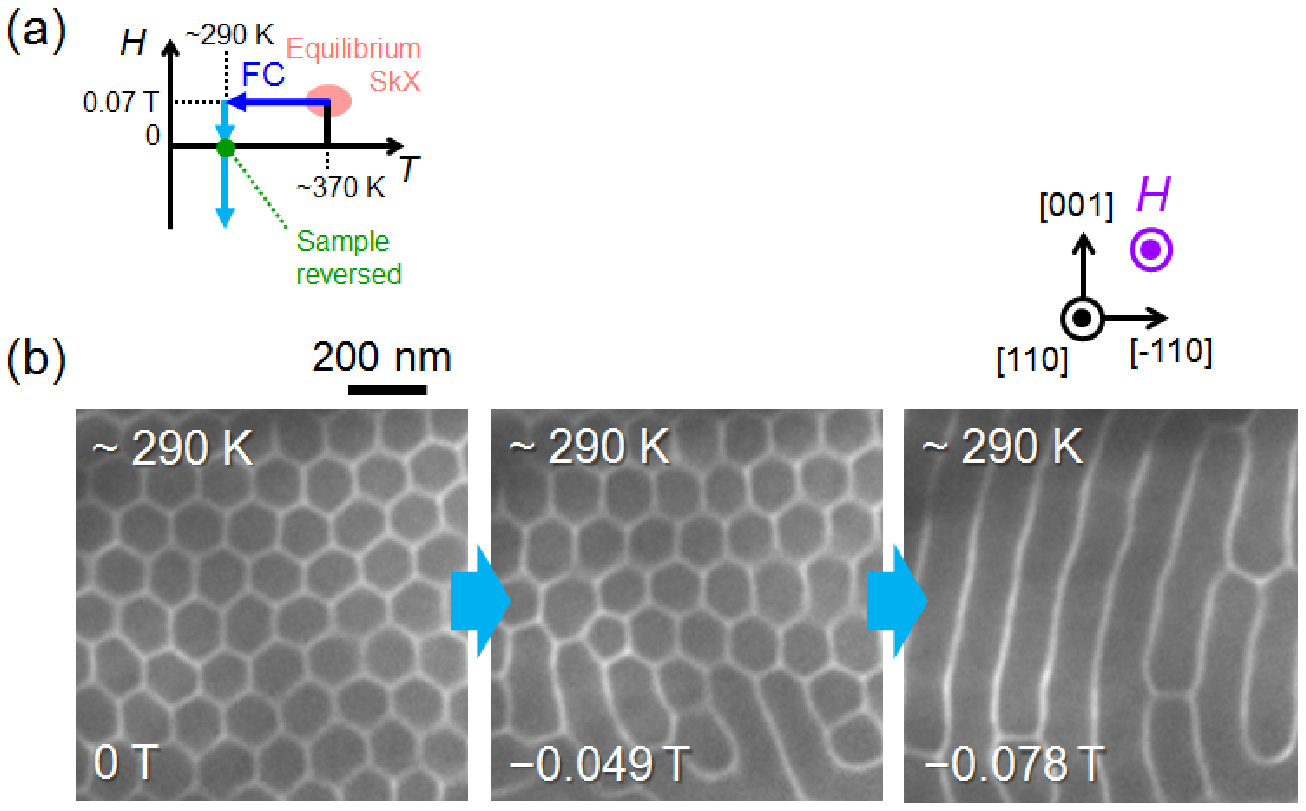}
\caption{(a) Schematic illustration of the measurement process for panel (b). 
After the FC via the equilibrium SkX phase to 290 K and the subsequent field-decreasing process to zero field, the sample was taken out of the electron microscope, turned upside down, and installed into the electron microscope again. Subsequently, the magnetic field was applied along the same (positive) direction for the electron microscope as before, which was the opposite (negative) direction for the sample. 
(b) Under-focused LTEM images on (110) plane at 290 K and selected magnetic fields in the field-sweeping process from zero field to negative fields, and after the above sample reversing procedure. 
The defocus values of the LTEM images were set to be $-$288 $\mu$m for 0 T and $-$0.049 T, and $-$384 $\mu$m for $-$0.078 T, respectively.
}
\end{center}
\end{figure}
%%%%%%%%%%%%%%%%%%%  FIG   %%%%%%%%%%%%%